\documentclass[
 reprint,
 superscriptaddress,
 showpacs,
 amsmath,amssymb,
 aps,
 prb,
]{revtex4-1}

\usepackage{graphicx}
\usepackage{dcolumn}
\usepackage{bm}
\usepackage{multirow}

\begin{document}


\title{Computational study of heavy group IV elements (Ge, Sn, Pb) triangular lattice atomic layers on SiC(0001) surface}

\author{Anton Visikovskiy}
 \altaffiliation{Department of Applied Quantum Physics and Nuclear Engineering, Faculty of Engineering, Kyushu University, 744 Motooka, Nishi-ku, Fukuoka-shi, Fukuoka 819-0395, Japan}
 \email{anton{\_}v@nucl.kyushu-u.ac.jp}
\author{Shingo Hayashi}%
\affiliation{%
 Department of Applied Quantum Physics and Nuclear Engineering, Faculty of Engineering, Kyushu University, 744 Motooka, Nishi-ku, Fukuoka-shi, Fukuoka 819-0395, Japan}%
\author{Takashi Kajiwara}%
\affiliation{%
 Department of Applied Quantum Physics and Nuclear Engineering, Faculty of Engineering, Kyushu University, 744 Motooka, Nishi-ku, Fukuoka-shi, Fukuoka 819-0395, Japan}%
\author{Fumio Komori}%
\affiliation{%
 Institute of Solid State Physics, University of Tokyo, Kashiwanoha, Kashiwa-shi, Chiba 277-8581, Japan}%
\author{Koichiro Yaji}%
\affiliation{Institute of Solid State Physics, University of Tokyo, Kashiwanoha, Kashiwa-shi, Chiba 277-8581, Japan}%
\author{Satoru Tanaka}%
\affiliation{%
 Department of Applied Quantum Physics and Nuclear Engineering, Faculty of Engineering, Kyushu University, 744 Motooka, Nishi-ku, Fukuoka-shi, Fukuoka 819-0395, Japan}%

\date{\today}

\begin{abstract}
Group IV heavy elements atomic layers are expected to show an interesting physical properties due to their large spin-orbit coupling (SOC). Using density functional theory (DFT) calculations with/without SOC we investigate the variation of group IV heavy elements overlayers, namely dense triangular lattice atomic layers (TLAL) on the surface of SiC(0001) semiconductor. The possibility of such layers formation and  their properties have not been addressed before. Here we show, that these layers may indeed be stable and, owing to peculiar bonding configuration, exhibit robust Dirac-like energy bands originating from $p_x+p_y$ orbitals and localized mostly within the layer, and $p_z$ band localized outside the layer and interacting with SiC substrate. We found that a $T_1$ adsorption site is most favorable for such TLAL structure and this results in an unusual SOC-induced spin polarization of the states around $\bar{K}$ points of Brillouin zone, namely the coexistence of Rashba- and Zeeman-like spin polarization of different states. We explain this phenomena in terms of symmetry of partial electronic density rather than symmetry of atomic structure.     
\end{abstract}

\maketitle


\section{\label{sec:intro}Introduction}
\begin{figure*}[ht]
    \centering
    \includegraphics[width=0.8\textwidth]{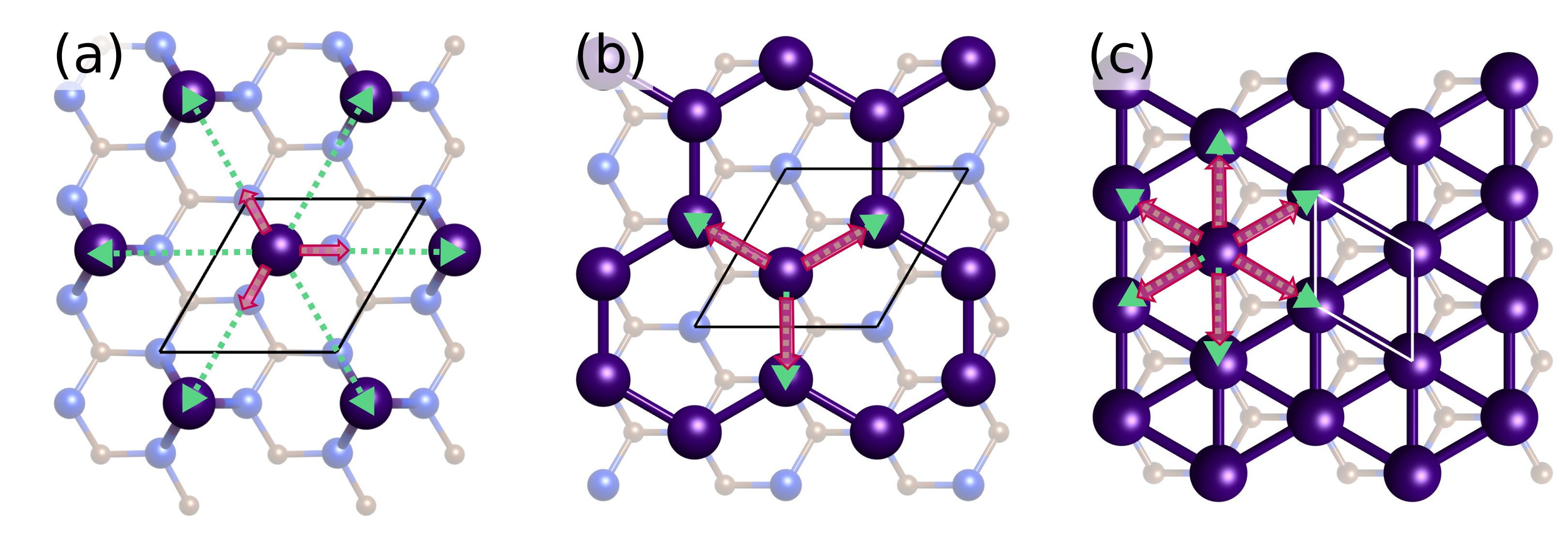}
    \caption{Adsorption of metal atoms on hexagonal surface of SiC(0001) with increasing atom density (red arrows indicate bonding and green arrows -- location of nearest neighbour adsorbate atoms): (a)~$\sqrt{3}\times\sqrt{3}$ sparse triangular adatom structure ($1/3$~ML), (b)~honeycomb graphene-like structure ($2/3$~ML), and (c)~dense triangular $(1\times1)$ structure.}
    \label{fig:triang}
\end{figure*}

Two-dimensional (2D) metal overlayers on semiconductor surfaces are excellent model systems for studying various quantum mechanics phenomena. Specifically, group IV heavy elements (Ge, Sn, Pb) triangular lattices on hexagonal semiconductor surfaces, especially Si(111) and Ge(111), are a long time study objects of surface scientists\cite{Sn-Si2,Sn-Ge1,Sn-Si1,Sn-Ge2,all-Si,cdw}. Surface charge density waves (CDW) were first observed by scanning tunneling microscopy (STM) in $(\sqrt{3}\times\sqrt{3})$-Pb overlayer on Ge(111)\cite{cdw}. While by simple electron count $(\sqrt{3}\times\sqrt{3})$ adatom overlayers of group~IV elements should be metallic, many of them exhibit insulating properties. This was explained by strong electron correlation and Mott- or Slater-type metal--insulator transition\cite{Sn-Si1,Sn-Ge2,all-Si}. The particular type of such transition is still under debate. As triangular arrangement of magnetic moments results in frustration, spin liquid phase has been expected to show up in these systems\cite{spin_liquid}. Experimentally, however, the spontaneous magnetic ordering has been observed\cite{mag_order}. Also, as heavy elements exhibit strong spin-orbit coupling (SOC), metallic overlayers show significant Rashba spin-polarization of electronic states, tunable both by choice of suitable substrate and by applying external gate voltage\cite{tune_Rashba}, making these structures advantageous in terms of application in spintronic devices. Typical $(\sqrt{3}\times\sqrt{3})$ adatom arrangement, however, is rather sparse (see Fig.~\ref{fig:triang}(a)). The adatoms are separated by significant distance ($5.33-5.65$~\AA), so no direct in-plane bonding exists in these structures. The electronic properties are governed by large-distance electron hopping and interaction via substrate. If the atomic density of such overlayer is increased by adding one more metal atom, this will result in honeycomb structure (Fig.~\ref{fig:triang}(b)). In honeycomb arrangement the atoms do interact directly, forming in-plane bonds. Honeycomb lattice results in formation of Dirac cone-like electron dispersion exhibiting many exotic properties similar to those of graphene and beyond. Group~IV elements honeycomb overlayers such as germanene \cite{germanene1,germanene2}, stanene\cite{stanene1,stanene2,stanene3}, and plumbene\cite{plumbene1,plumbebe2} have boomed recently in number of experimental and theoretical works. Yet, these are still difficult to grow on semiconductor surfaces and were mostly synthesized on metals, making it less suitable for direct device applications. Further increase in atomic density by adding one more atom will result again in triangular structure (Fig.~\ref{fig:triang}(c)). This, however, is rather different from $(\sqrt{3}\times\sqrt{3})$ phase mentioned previously. In such a dense triangular lattice atoms are very close to each other ($3.08-3.26$~\AA), making direct in-plane bonding possible. The bonding configuration in such system is very peculiar. Each metal atom has six nearest neighbours, while typically group IV elements tend to form three or four covalent bonds, depending on orbital hybridization. This may result in layers with new interesting physics and useful properties. While undoubtedly interesting, the properties of such systems are mostly unexplored, mainly due to the lack of experimentally observed structures of this type. The only reports of closely related systems are Si(111)-$(1\times1)$-Tl\cite{thallium1,thallium2} and high-density Si(111)- and Ge(111)-$(\sqrt{3}\times\sqrt{3})$-Pb\cite{pb_dense1,pb_dense2,pb_dense3,pb_dense4}. In thallium system adatoms sit at $T_4$ site of bulk-terminated Si(111) surface\cite{thallium1}. Due to the symmetry of the atomic structure ($C_3$ symmetry at $\bar{K}$ point of Brillouin zone), Tl system exhibit so-called Zeeman-type of spin-splitting of states at the $\bar{K}$ point with spin direction normal to the surface\cite{tl_Rashba,tl_Rashba_bi}. In the Pb system the one of the atom sits at $H_3$ site (though \textit{ab initio} calculations shows that system with adatom at $T_4$ is just marginally higher in energy) and other three atoms are at slightly displaced locations from main symmetric adsorption sites\cite{pb_dense2,yaji1,yaji2}. In this structure Zeeman-like splitting has been also observed at $\bar{K}_{\sqrt{3}}$ point, as well as small Rashba-type splitting at time-reversal-invariant momenta (TRIM) points ($\bar{M}$ and $\bar{M}_{\sqrt{3}}$)\cite{pb_dense3}. The conventional Rashba type splitting is described by Rashba Hamiltonian $H_{R}(\mathbf{k})=\alpha_{R}(k_{x}\sigma_{y}-k_{y}\sigma_{x}))$, where $\sigma_{x,y}$ and $k_{x,y}$ are Pauli matrices and in-plane wave-vector component respectively, and $\alpha_{R}$ is a Rashba parameter\cite{rashba}. It is observed only in TRIM points of the Brillouin zone and results in in-plane spin vortexes. Some unconventional Rashba effects has been also reported to occur at non-TRIM points, such as $\bar{K}$ point, for systems like Si(111)-$(\sqrt{3}\times\sqrt{3})$-Bi with $C_{3v}$ symmetry at $\bar{K}$\cite{Sakamoto_bi}. While, not being classical Rashba effect, this spin-polarization is well described by Rashba-like Hamiltonian\cite{Sakamoto_bi,tl_Rashba_bi}.  

The lack of experimental evidences of dense triangular lattice atomic layers (TLALs) of group~IV is due to the fact that most of up to date the adsorption experiments has been carried out on Si(111) or Ge(111) substrates, which unit cell is too large for stable TLAL formation. SiC(0001) surface, however, is much more suitable candidate for TLALs. In fact we have experimentally observed Sn TLAL structure formation at graphene/SiC(0001) interface by Sn atoms intercalation\cite{Hayashi}. In the present paper we show that dense triangular overlayers with symmetric adatoms locations are possible on bare SiC(0001) surface and discuss their peculiar electronic structure.

\section{\label{sec:comp}Computation details}
The calculations have been performed using Vienna \textit{ab initio} simulation package (VASP)\cite{vasp} with PAW-type pseudopotentials\cite{paw1,paw2}. General gradient approximation in Perdew-Burke-Ernzerhof (PBE)\cite{pbe} formulation has been used for exchange and correlation. Plane wave cut-off energy was set to 500~eV. In structure optimization $\Gamma$-point centered regular $7\times7\times1$ and $5\times5\times1$ k-space samplings were used for $(1\times1)$ and $(\sqrt{3}\times\sqrt{3})$ structures respectively. For accurate ground state energy and electronic density calculations k-space mesh density has been increased to $13\times13\times1$ and $7\times7\times1$, respectively. In most cases structure optimization has been performed without SOC included to save time, though several tests were performed to ensure that inclusion of SOC does not have significant influence on the resulting structure or total energy difference when comparing different structures. The structures were modelled as a slabs of 6H-SiC(0001) with six SiC bilayers, back surface saturated with hydrogen atoms, top surface covered by metal overlayer, and vacuum layer of 10~\AA{} separating slab periodic copies. The structure optimization has been performed until maximum force is less than 0.005~eV/\AA{}.

\section{\label{sec:result}Results and discussion}
\subsection{\label{subsec:site}$(1\times1)$ adsorption site}
\begin{figure}[ht]
    \centering
    \includegraphics[width=1.0\columnwidth]{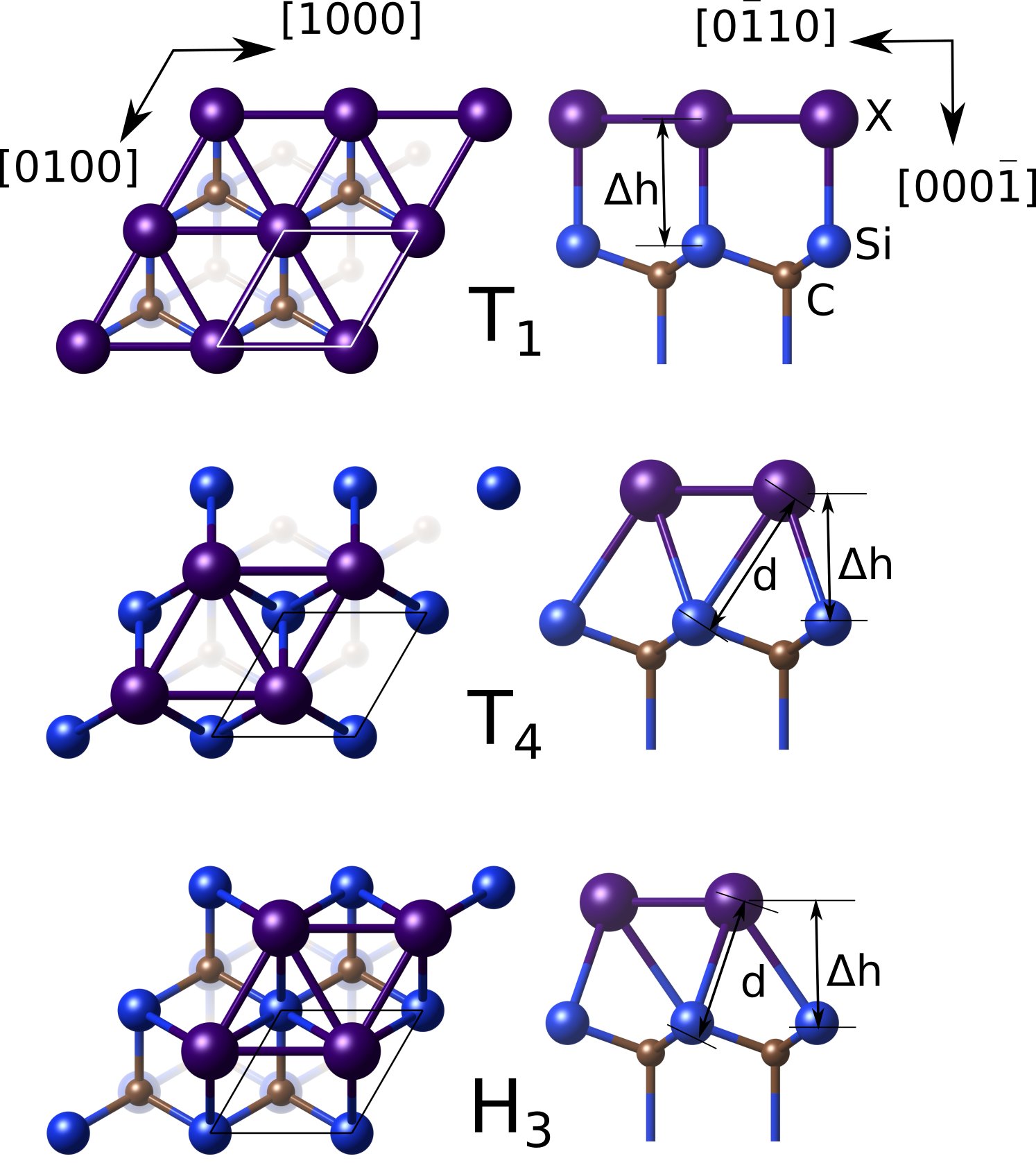}
    \caption{SiC(0001)-$(1\times1)$-X atomic models. Three adsoption sites are possible for simple 1~ML overlayer: on-top or $T_1$ site, tetrahedral $T_4$ site, and hollow $H_3$ site}
    \label{fig:site}
\end{figure}

As the model for dense TLALs on SiC(0001) was initially guessed to be $(1\times1)$, which is in agreement with our previous experimental results\cite{Hayashi}, first we investigate different models of $(1\times1)$-X overlayers (where X is Ge, Sn, or Pb atoms) on SiC(0001). As $(1\times1)$ superstructure is rather small, only limited number of structural models, distinct only by X atom adsorption site, are reasonable. Three main adsorption sites exist on hexagonal SiC(0001) surface, namely $T_1$ or on-top site, and tetrahedrally coordinated $T_4$ and $H_3$ sites (Fig.~\ref{fig:site}). Table~\ref{tab:site} shows the calculated relative total energies and structural parameters  for these models. For all three elements on-top $T_1$ site is found to be most stable by quite substantial energy difference. This is in contrast with Si(111)-$(1\times1)$-Tl system, where Tl atoms theoretically and experimentally are found to occupy $T_4$ site\cite{thallium1,thallium2}. This difference has significant consequences for electronic structure as will be shown below. Interestingly interlayer distances in all three adsorption configurations are very close. This is because in $(1\times1)$ structure, tetrahedrally coordinated adsorbate atoms (in $T_4$ or $H_3$ sites) have to share bonds with substrate top Si atoms, making these bonds weaker and much longer compared to strong bonds in on-top configuration, where each adsorbate atoms has one individual bond with substrate Si. Experimentally the negligible difference in interlayer distances could make it difficult to distinguish between model when layer distance sensitive only methods are used (such as X-ray truncation rod scattering analysis used in our previous experimental work \cite{Hayashi}).

\begin{table}[]
    \centering
    {\def\arraystretch{1.5}
    \begin{tabular}{c|c|c|c|c}
    \hline\hline
 & & $T_1$ site & $T_4$ site & $H_3$ site \\\hline
 & Ge & $E_{ref}^{Ge}$  & $E_{ref}^{Ge}$+0.67  & $E_{ref}^{Ge}$+0.43  \\
$E_{tot}$, eV & Sn & $E_{ref}^{Sn}$  & $E_{ref}^{Sn}$+0.51  & $E_{ref}^{Sn}$+0.34  \\
 & Pb & $E_{ref}^{Pb}$  & $E_{ref}^{Pb}$+0.40  & $E_{ref}^{Pb}$+0.20  \\\hline
 & Ge & 2.51 & 2.40 & 2.47  \\
 $\Delta h$, \AA{} & Sn & 2.71 & 2.65 & 2.59  \\
 & Pb & 2.81 & 2.79 & 2.71  \\\hline
 & Ge & 2.51 & 3.05 & 2.99  \\
 $d$, \AA{} & Sn & 2.71 & 3.20 & 3.15  \\
 & Pb & 2.81 & 2.79 & 2.71  \\
    \hline\hline
    \end{tabular}}
    \caption{Results of structure optimization and total energy calculations of SiC(0001)-$(1\times1)$-X structures with different adsorption sites.}
    \label{tab:site}
\end{table}

\subsection{\label{subsec:energy}Stability of $(1\times1)$ overlayer}
\begin{figure}[ht]
    \centering
    \includegraphics[width=0.7\columnwidth]{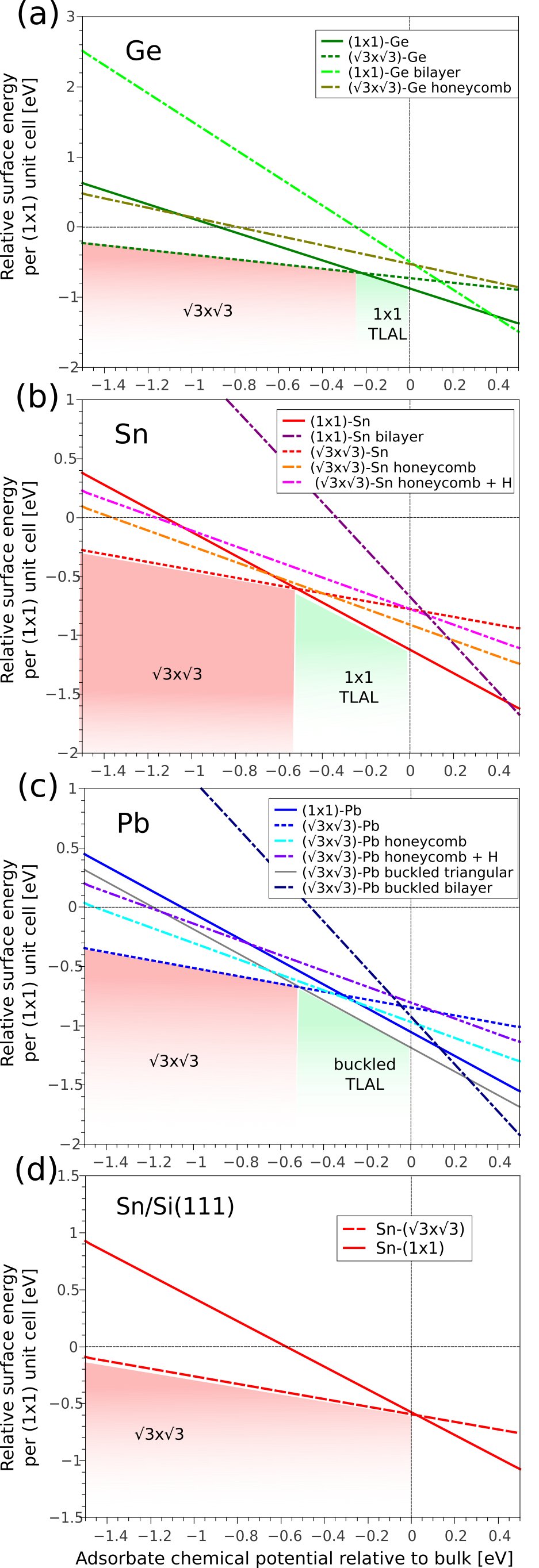}
    \caption{(a)-(c)Calculated relative surface energy of various surface structures of adsorbate atoms on SiC(0001), (d) relative surface energy of Sn layers on Si(111) is given for reference.}
    \label{fig:energy}
\end{figure}

As mentioned before, one of the main reasons the properties of $(1\times1)$ group~IV elements triangular overlayer has not been addressed before is that, with rare exceptions, no such overlayer has been observed so far. Here we address the principal possibility of that kind of structure to exist on SiC(0001). After we have confirmed that among different models, the one with $T_1$ on-top adsorption site is most stable, we may compare the relative  surface energy of this structure to more conventional reconstructions observed and/or proposed. That is $(\sqrt{3}\times\sqrt{3})$ $1/3$~ML adatom structure, $(\sqrt{3}\times\sqrt{3})$ honeycomb layer with and without hydrogen termination of the remaining dangling bonds similar to those of bismuthen reported in Ref.~\onlinecite{bismuthene}, and $(1\times1)$ bilayer. The $1.25$-ML $(\sqrt{3}\times\sqrt{3})$ triangular arrangement similar to that of Pb on Si(111) or Ge(111) surface\cite{pb_dense1,pb_dense2,pb_dense3,pb_dense4,yaji1,yaji2} is not considered here due to large lattice mismatch between such arrangement on much more compact SiC(0001) and equilibrium unit cell of artificial standalone TLAL (2.78, 3.16, and 3.29~\AA{} for Ge, Sn and Pb respectively). As the different surface structures are of different adsorbate atoms coverage, one cannot compare the calculated total energies directly, but  have to build dependence on adsorbate chemical potential $\mu_X$. Relative surface energy is calculated in the following way:
\begin{equation}
    \gamma=\frac{1}{A}(E_{tot}-E_{SiC}-N_{X}\mu_{X}),
\end{equation}
where $\gamma$ is relative surface energy per $(1\times1)$ unit cell, $A$ -- unit cell surface area (in terms $1\times1$ cells), $E_{tot}$ -- total energy of a system, $E_{SiC}$ -- energy of ideal bulk terminated SiC slab without adsorbates, $N_X$ -- number of adsorbate atoms, and $\mu_X$ -- chemical potential of atom X. The results are shown in Fig.~\ref{fig:energy} using the chemical potential scale relative to the corresponding bulk values (calculated separately). Thus, the positive values on horizontal axis mean that three-dimensional (3D) growth of bulk material is more preferable, while negative values indicate region where 2D layer is more stable. As seen from the Fig.~\ref{fig:energy} all three elements show the existence of stability region for $(1\times1)$ reconstruction in the adsorbate rich conditions (the buckling case of Pb will be discussed below). In the adsorbate poor region normal $(\sqrt{3}\times\sqrt{3})$ adatom structure prevails. Note, that all other simple reconstructions show higher relative surface energies and should be considered unstable. For comparison in Fig.~\ref{fig:energy}(d) similar calculations for Sn/Si(111) surface are shown. In this case the $(1\times1)$ structure is unstable. The reason for this is substantially larger surface unit cell of 3.84~\AA{} of Si(111) compared to 3.08~\AA{} of SiC(0001). At such long distance the direct in-plane bonding between adsorbate atoms seems to be unfavorable with shorter equilibrium unit cells of free standing TLALs. This is one of the reasons why these structures have not been observed before, as SiC(0001) surface is much less studied experimentally than Si(111).

\subsection{\label{subsec:buckling}Buckling of triangular overlayer}
\begin{figure*}[ht]
    \centering
    \includegraphics[width=0.9\textwidth]{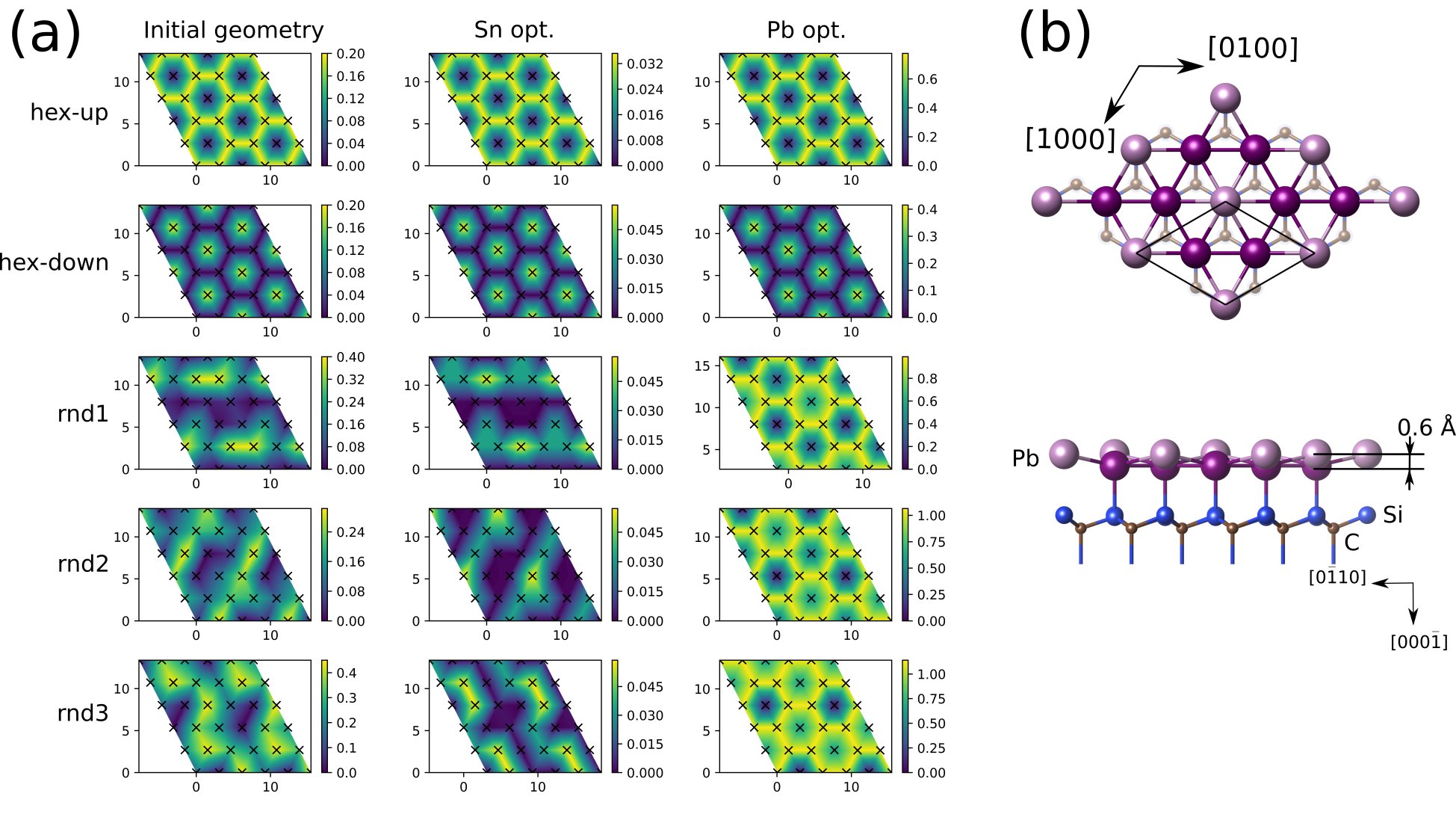}
    \caption{(a) Height maps of TLAL layers optimized in $(3\times3)$ supercell conditions (each map shows four $(3\times3)$ cells, $z$ axis in calculations was pointing downwards, so smaller numbers actually mean higher adsorbate atoms). The initial vertical position of atoms is perturbed in a regular or random manner as shown in left column (details in text). The optimized structures are shown for Sn layer in the middle column and for Pb in the right column. Crosses show the location of adsorbate atoms. (b) Schematic view of optimized buckled structure of Pb overlayer, $(\sqrt{3}\times\sqrt{3})$ cell in shown, purple balls are lower Pb atoms forming honeycomb, and light purple ball is higher Pb atom.}
    \label{fig:buckl}
\end{figure*}

As mentioned in the previous section the equilibrium cell size of free-standing triangular metallic layer is slightly different to SiC(0001) surface unit cell size. While for Ge case the equilibrium size is smaller, so $(1\times1)$ layer will be slightly stretched and planar geometry is naturally expected, this cannot be said about Sn and Pb case, which both have slightly larger equilibrium cell size ($3.16$ and $3.29$~\AA{} respectively). The compressive stress on a layer may lead to buckling and, as a result, increase of the actual periodicity of the overlayer structure. Thus, to check this possibility we performed calculations with large unit cells of $(3\times3)$ and $(6\times6)$ with regularly and randomly displaced metal atoms in vertical direction to reduce any artificially induced symmetry constrains on a system. Then, structure optimization calculations has been performed. These were repeated for several regular and random initial configurations. Fig.~\ref{fig:buckl} represent a height map of adsorbate atoms before and after structure optimization. In Table~\ref{tab:relax} the energetics of different structures are compared. It is clear that the structures with higher than $(1\times1)$ periodicity are formed. In the case of Sn atoms, the overlayer corrugation after structure optimization is rather small ($<0.1$~\AA{}) and in most cases retain the similar height map as original structure, which is indicative for artifact of structure optimization routine (such as meeting convergence criteria before perfect arrangement is formed). Also the energy differences between ideal $(1\times1)$ and larger, slightly buckled structures are negligible. Thus, we may safely assume that Sn forms non-buckled $(1\times1)$ overlayer on SiC(0001). On the other hand, Pb layer optimization shows significantly buckled structure. The basic periodicity of resultant overlayer from almost all trial configurations is close to $(\sqrt{3}\times\sqrt{3})$. The overlayer is represented by honeycomb planar layer plus the atom in the center of the hexagon with slightly higher vertical position ($+0.6$~\AA{}) as shown in Fig.~\ref{fig:buckl}. As seen from the Table.~\ref{tab:relax} this reconstruction is noticeably lower in energy than $(1\times1)$, making the stability region of this dense overlayer even wider.

\begin{table}[]
    \centering
    {\def\arraystretch{1.5}
    \begin{tabular}{c|c|c|c|c}
        \hline\hline
                     Element &  Trial config. &    & Before opt. & After opt. \\\hline
                             & hex-up   & $\delta h$, \AA{} & 0.2  & 0.03  \\
                             &                           & $\gamma$, eV      &  --- & $\gamma_{1x1}+0.05$ \\\cline{2-5}
                             & hex-down & $\delta h$, \AA{} & 0.2  & 0.05  \\
                             &                           & $\gamma$, eV      & ---  & $\gamma_{1x1}+0.04$  \\\cline{2-5}
                          Sn & rnd1     & $\delta h$, \AA{} & 0.4  & 0.05  \\
                             &                           & $\gamma$, eV      &  --- & $\gamma_{1x1}+0.03$  \\\cline{2-5}
                             & rnd2     & $\delta h$, \AA{} & 0.3  & 0.05  \\
                             &                           & $\gamma$, eV      &  --- & $\gamma_{1x1}+0.03$  \\\cline{2-5}
                             & rnd3     & $\delta h$, \AA{} & 0.45 & 0.05  \\
                             &                           & $\gamma$, eV      &  --- & $\gamma_{1x1}+0.03$  \\\hline
                             & hex-up   & $\delta h$, \AA{} & 0.2  & 0.7  \\
                             &                           & $\gamma$, eV      &  --- & $\gamma_{1x1}-0.13$ \\\cline{2-5}
                             & hex-down & $\delta h$, \AA{} & 0.2  & 0.4  \\
                             &                           & $\gamma$, eV      &  --- & $\gamma_{1x1}-0.02$  \\\cline{2-5}
                         Pb  & rnd1     & $\delta h$, \AA{} & 0.4  & 0.9  \\
                             &                           & $\gamma$, eV      &  --- & $\gamma_{1x1}-0.13$  \\\cline{2-5}
                             & rnd2     & $\delta h$, \AA{} & 0.3  & 0.9  \\
                             &                           & $\gamma$, eV      &  --- & $\gamma_{1x1}-0.13$  \\\cline{2-5}
                             & rnd3     & $\delta h$, \AA{} & 0.45 & 1.0  \\
                             &                           & $\gamma$, eV      &  --- & $\gamma_{1x1}-0.13$  \\\hline\hline
    \end{tabular}}
    \caption{Results of structure optimization of larger $(3\times3)$ unit cell with regular or random overlayer buckling. 'hex-up' structure correspond to $(\sqrt{3}\times\sqrt{3})$ honeycomb planar structure plus one atom higher above surface, 'hex-down' correspond to $(\sqrt{3}\times\sqrt{3})$ honeycomb planar structure plus one atom lower above surface, 'rnd1', 'rnd2','rnd3' are models with randomly displaced atoms in vertical direction.}
    \label{tab:relax}
\end{table}

\subsection{\label{subsec:band}Band structure}
\begin{figure*}[ht]
    \centering
    \includegraphics[width=1.0\textwidth]{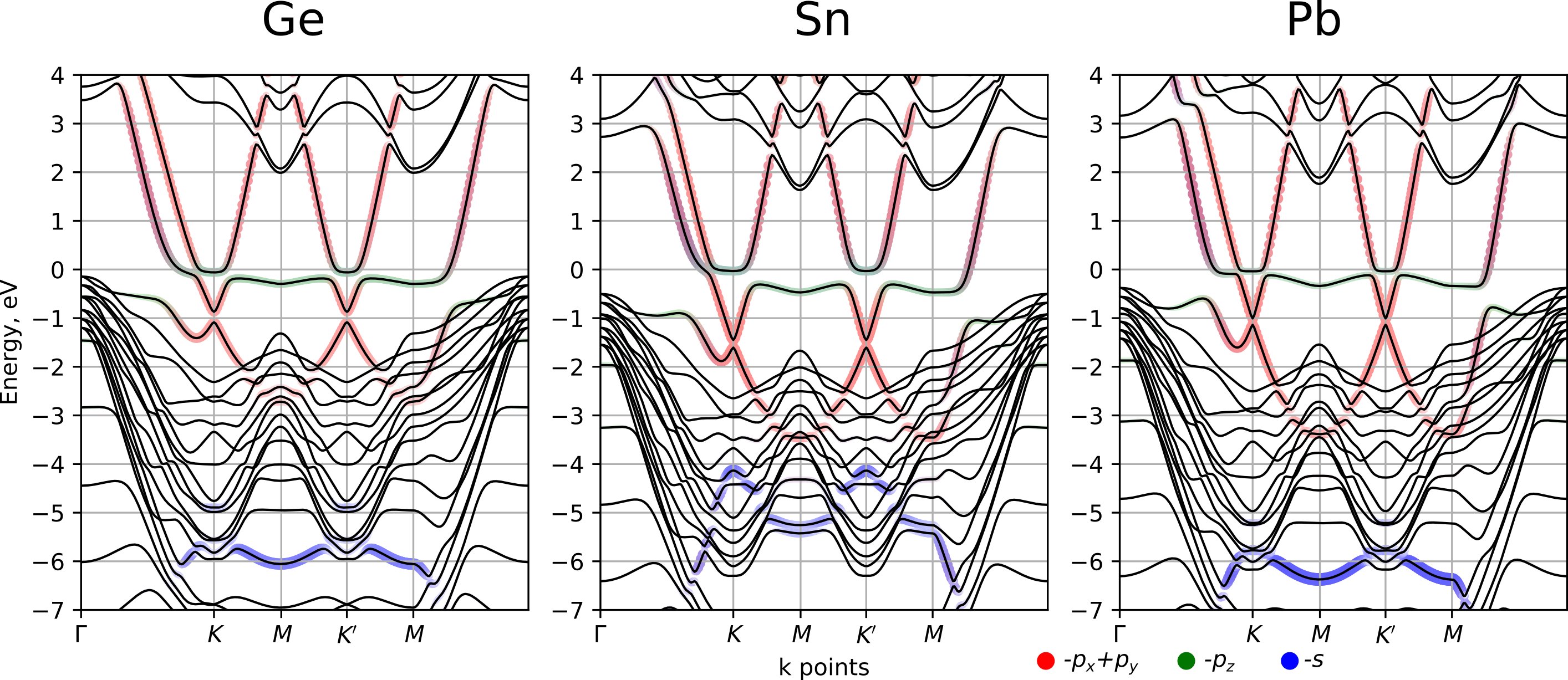}
    \caption{Results of band structure calculations for Ge, Sn, and Pb $(1\times1)$ overlayer on SiC(0001). The colored bands are derived from $s$-, $p_z$-, and $(p_x+p_y)$-orbitals as was found using orbital decomposition. Note, that Dirac-like bands have primary $(p_x+p_y)$ contribution. Also flattish band around Fermi level has mainly $p_z$ character with some $s$ orbital mixture (bluish shade). Also for Pb the bands structure of unbuckled $(1\times1)$ overlayer is shown rather than optimized buckled $\sqrt{3}\times\sqrt{3}$ one. We show in Supplementary\cite{suppl}, that unfolding buckled overlayer band structure onto $(1\times1)$ Brillouin zone results in almost identical bands as unbuckled structure.}
    \label{fig:bands}
\end{figure*}

\begin{figure*}[ht]
    \centering
    \includegraphics[width=1.0\textwidth]{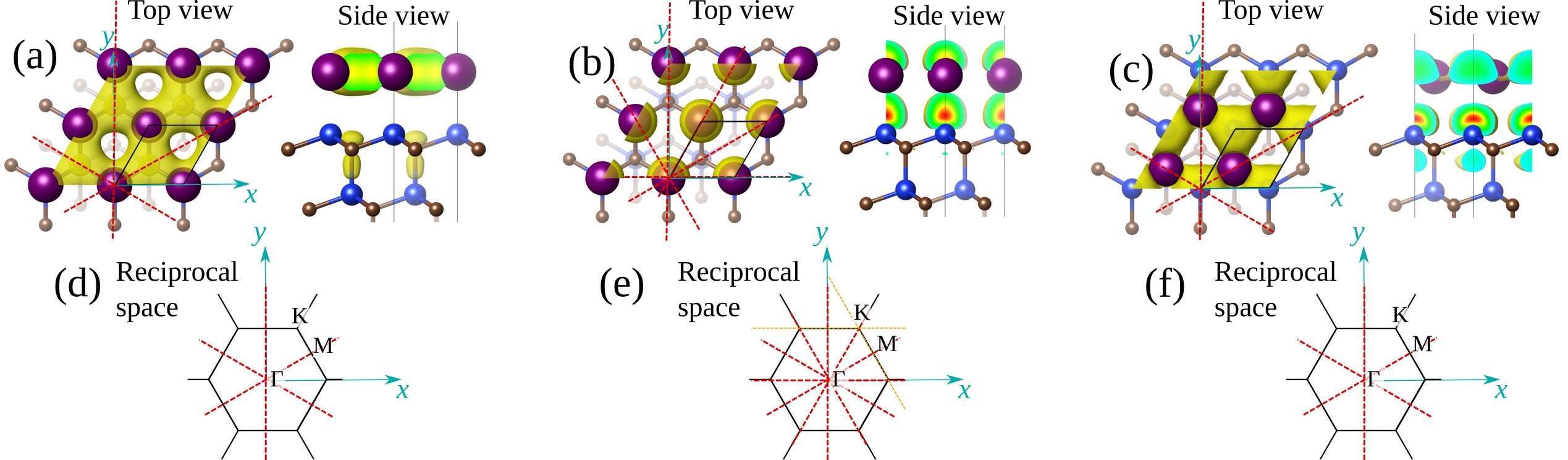}
    \caption{Calculated partial electronic density distribution for states of interest of SiC(0001)-$(1\times1)$-Sn system. (a) Partial density of a state originating primary of $p_x+p_y$ states exhibiting $p3m1$ symmetry. Notice one mirror plane shown with dashed line. (b) Partial density of state originating from $p_z+s$ state. Notice six mirror planes and $p6mm$ symmetry. (c) Density of $p_z+s$ state for artificial structure with Sn atom located at $T_4$ site (similar to Si(111)-$(1\times1)$-Tl structure). Notice that unlike previous case, the density exhibit one mirror plane and $p3m1$ symmetry, thus no Rashba effect is observed for such structure. (d)-(f) Schematic representation of Brillouin zone and corresponding mirror planes showing the local symmetry at $\bar{K}$ point to be $C_3$ in (d) and (f), and $C_{3v}$ in (e).}
    \label{fig:density}
\end{figure*}

The typical calculated band structure without SOC is shown in Fig.~\ref{fig:bands}. All elements exhibit qualitatively similar band structure with some variation of the energy of particular bands and values of gaps. The two main features of band structure are Dirac-cone-like dispersion at $\bar{K}$ and $\bar{K'}$ points around $-1.0~-1.5$~eV below Fermi level and much flatter band around Fermi level. Interestingly the Dirac-like dispersion feature originate totally from $p_x$ and $p_y$ orbitals as seen from orbital projection calculation rather than $p_z$ orbital like in case of most honeycomb lattices. Electronic density of these bands is fully localized inside the TLAL, which makes it a perfect 2D metal (see Fig.~\ref{fig:density}(a)). The branches of Dirac-like dispersion exhibit a small gap (220~meV, 160~meV, and 130~meV for Ge, Sn, and Pb respectively). The origin of this gap is similar to the one discussed in Ref.~\onlinecite{sp-tight-binding}, namely the braking of space inversion symmetry, which allows electron hopping from $p_z$ orbital (hybridized with substrate dangling bonds) states to $p_x$ and $p_y$ states. In the band structure of artificial free-standing triangular layer these gaps are absent (see Supplementary\cite{suppl} Fig.~S1). The $p_z$ orbital, on the other hand, hybridize slightly with $s$ orbital and top layer Si dangling bonds and form flattish band near the Fermi level.

As discussed in previous section Pb TLAL is buckled and has $(\sqrt{3}\times\sqrt{3})$ periodicity rather than $(1\times1)$. This however does not influence bands structure to significant degree. If we perform unfolding procedure\cite{unfolding} of $(\sqrt{3}\times\sqrt{3})$ band structure onto $(1\times1)$ Brillouin zone, one can see that the bands are actually almost identical to those calculated for planar $(1\times1)$ layer. So, for band properties we restrict our further study to planar Pb layer. The full and unfolded band structure of buckled Pb layer can be found in Supplementary\cite{suppl}(Fig.~S2). 

The most interesting results are obtained in calculations including SOC. The typical band structure is shown in Fig.~\ref{fig:spin} (the element specific plots could be found in Supplementary\cite{suppl}(Fig.~S3)). The part of the $p_z$ band close to Fermi level shows Rashba-like spin-polarization split around $\bar{K}$ and $\bar{K'}$ points. The calculated spin texture shows characteristic vortex like structure, shown in Fig.~\ref{fig:spin}(c). It has to be noted that small normal spin component is still present in contrast with ideal classic Rashba-Bychkov effect. At the same time, Dirac-like states originating from $p_x$ and $p_y$ orbitals exhibit Zeeman-like spin-polarization with primarily normal spin components and vanishing in-plane component close to the same $\bar{K}$ and $\bar{K'}$ points (Fig.~\ref{fig:spin}(b,~d)) with substantial $\Delta E_Z$ of 120, 170, and 160~meV for Ge, Sn and Pb cases respectively.

\begin{figure*}[ht]
    \centering
    \includegraphics[width=1.0\textwidth]{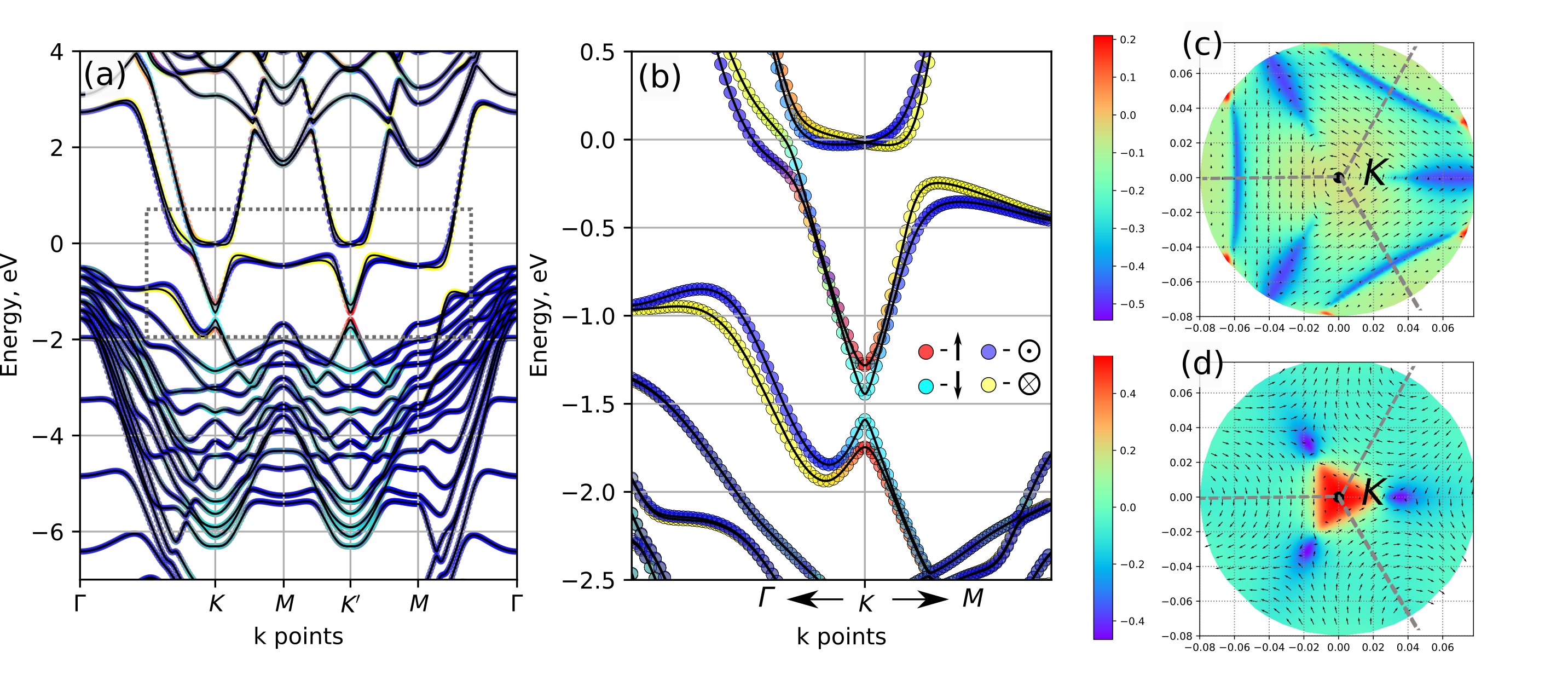}
    \caption{This is a characteristic result of band structure including SOC of group IV TLAL shown using Sn as an example. (a) The large portion of the Sn band structure including SOC. Color represents spin orientation as shown in the legend in the figure (b). Note, that only states shown in dashed rectangle are non-degenerate and exhibit spin polarization. (b) The magnified portion of the band structure around $\bar{K}$ point. Notice two types of spin polarization at the $\bar{K}$ point: Zeeman-like for cone-like states (with spin orientation normal to the surface), and Rashba-like for $p_z$ flattish states near Fermi level. (c) The spin texture around $\bar{K}$ point of the outer Rashba-type band laying near Fermi-level. The in-plane spin is illustrated by arrows, while color represent surface-normal component. (d) Same for the cone-like Zeeman-splitted band. Note the quite abrupt increase in spin z-component, while vanishing in-plane texture.}
    \label{fig:spin}
\end{figure*}

These results look very unusual. $\bar{K}$ and $\bar{K'}$ points are non-TRIM points of the Brillouin zone, so no classic Rashba effect is expected. It has been shown, however, that in triangular systems with $p31m$ atomic symmetry (such as Si(111)-$(\sqrt{3}\times\sqrt{3})$-Bi) the local symmetry at $\bar{K}$ points is $C_{3v}$ (Fig.~\ref{fig:density}(e)) and in-plane, vortex-like spin-polarization is possible and has formalism similar to Rashba effect (hence called unconventional Rashba effect)\cite{Sakamoto_bi,kp_rashba}. In systems with $p3m1$ symmetry, like Si(111)-$(1\times1)$-Tl, the local symmetry at $\bar{K}$ points is just $C_3$ (Fig.~\ref{fig:density}(d, f)) and in-plane spin components are forbidden. However, surface normal component exists owing to normal effective magnetic field due to atomic character of wavefunctions and their average non-zero angular momentum, resulting in Zeeman-like spin polarization\cite{tl_Rashba,tl_Rashba_bi}. In our case, the total atomic system symmetry is $p3m1$, but still, at the vicinity of $\bar{K}$ point, the both effects are observed simultaneously for different bands. Interestingly, the calculation of the same triangular layers residing on $T_4$ site produce a gap in $p_z$ split-band crossing point at $\bar{K}$ and Zeeman-like polarization also for $p_z$ states, similar to reported Si(111)-$(1\times1)$-Tl case\cite{tl_Rashba} (see Supplementary\cite{suppl} Fig.~S4). Hence, the optimal calculated $T_1$ site position of our TLAL on SiC(0001) must play a crucial role in creating a Rashba-like band splitting near Fermi level. Our suggestion is that not the symmetry of atomic structure exclusively, but the symmetry of partial electron density of electronic states play an important role in the type of spin-polarization observed near special points of the Brillouin zone at least at phenomenological viewpoint. The ideal triangular layer itself possesses $p6mm$ symmetry, which results in $C_{3v}$ local symmetry at $\bar{K}$ points (Fig.~\ref{fig:density}(e)). Hence, the unconventional Rashba-type state splitting should be observed in the case of inversion symmetry breaking. Indeed, if we model the single side hydrogenated TLAL layer (system which retains $p6mm$ symmetry) we may see the Rashba splitting in SOC calculation of the $p_z+s$ related states, while there are neither Rashba nor Zeeman spin-splitting in $p_x+p_y$ states at $\bar{K}$ points (see Supplementary Fig.~S5(c)). The partial charge density of all states is also $p6mm$ symmetric because there are no symmetry breaking perturbations (Fig.~S5(a)). However, if we model the TLAL layer on SiC bilayer, even at relatively large interlayer distances of $\sim6$~\AA{} the perturbation of substrate with lower symmetry is enough to rearrange partial electronic density of $p_x+p_y$ states to those shown in Fig.~\ref{fig:density}(a) with $p3m1$ symmetry (Fig.~S5(b)). The reason for that is metastable nature of $p_x+p_y$ states in high symmetry environment. As there are six nearest neighbors and just three electrons to share per atom. The configuration with lower symmetry, however, creates 3-fold symmetric partial electron density much better suitable for group IV atoms. The details of such symmetry transformation is beyond the scope of the present paper and will require deep theoretical investigation. Hence, $p_x+p_y$ state has $p3m1$ symmetry and $C_{3}$ symmetry at $\bar{K}$ point (Fig.~\ref{fig:density}(d)).  Inclusion of SOC leads to Zeeman-like splitting of these states. This phenomena would be universal for $p_x+p_y$ bands of TLAL on 3-fold symmetric substrate independent on the adsorption site. On the other hand, the states originating from $p_z$ orbital on SiC substrate couple with topmost Si dangling bonds. The degree of such coupling and prevailing character of symmetry of the resultant electron density determine which type of splitting will prevail. In the case of triangular layer adsorption on $T_4$ site, each $p_z$ orbital of metal atom will form $\pi$-bond with three nearest Si dangling bonds, resulting in prevailing $p3m1$ symmetry of the mixed state (Fig.~\ref{fig:density}(c)). So, the Zeeman-type spin-splitting is observed at $\bar{K}$ points (see Supplementary\cite{suppl} Fig.~S4). It has to be noted that some character or Rashba-effect is still persis (so-called Rashba-Zeeman subband feature, proposed previously for Rashba materials in the external normal magnetic field)\cite{RZ_subband1,RZ_subband2}, as one can see that the minima of parabolic bands are still shifted in k-space from $\bar{K}$ points (this may be also the reason of the flattening of the bottoms of parabolic bands at $\bar{K}$ points observed in Si(111)-$(1\times1)$-Tl system\cite{tl_Rashba}). In the case of $T_1$ adsorption, the $p_z$ orbital of metal atoms are sitting directly on top of Si dangling bonds. This configuration conserves $p6mm$ symmetry of partial charge density to a high degree (Fig.~\ref{fig:density}(b)), and, as a result, Rashba-like splitting is still prevail, with band crossing (or a negligibly small gap) at $\bar{K}$ points. We have to note also, that existence of both Rashba- and Zeeman-type splittings have been observed experimentally by spin angle-resolved photoemission spectroscopy for Sn TLAL at graphene/SiC(0001) interface\cite{yaji}. 

\section{\label{sec:Conclusions}Conclusions}
In conclusion, we have investigated by means of DFT calculations the possibility of formation dense  triangular overlayer of Ge, Sn, or Pb on SiC(0001) surface and its electronic structure. We have found that for Ge and Sn $(1\times1)$ simple structure with adsorbate atom at $T_1$ position is stable in the adsorbate rich conditions. For Pb, due to the larger atomic radius, the buckled structure is stable and has $(\sqrt{3}\times\sqrt{3})$ periodicity which, however, does not reflect much on electronic properties. The band structure of TLAL exhibits characteristic cone-like feature at $\bar{K}$ points below the Fermi level originating mainly from $p_x+p_y$ orbitals of adsorbate atom. These bands show Zeeman-like spin-splitting upon inclusion of SOC into account. Another band close to Fermi level and originating primarily from $p_z$ orbital shows Rashba-like spin polarization and spin vortices around non-TRIM $\bar{K}$ point. This different behavior of bands is attributed to different symmetry of partial electronic density and is an important insight to understand the SOC-induced mechanism of bands spin-polarization.  

\onecolumngrid
\clearpage
\begin{center}
\textbf{\LARGE Supplemental Materials: Computational study of heavy group IV elements (Ge, Sn, Pb) triangular lattice atomic layers on SiC(0001) surface}
\end{center}
\setcounter{equation}{0}
\setcounter{figure}{0}
\setcounter{table}{0}
\setcounter{page}{1}
\makeatletter
\renewcommand{\theequation}{S\arabic{equation}}
\renewcommand{\thefigure}{S\arabic{figure}}
\renewcommand{\bibnumfmt}[1]{[S#1]}
\renewcommand{\citenumfont}[1]{S#1}

\begin{figure*}[ht]
    \centering
    \includegraphics[width=0.45\textwidth]{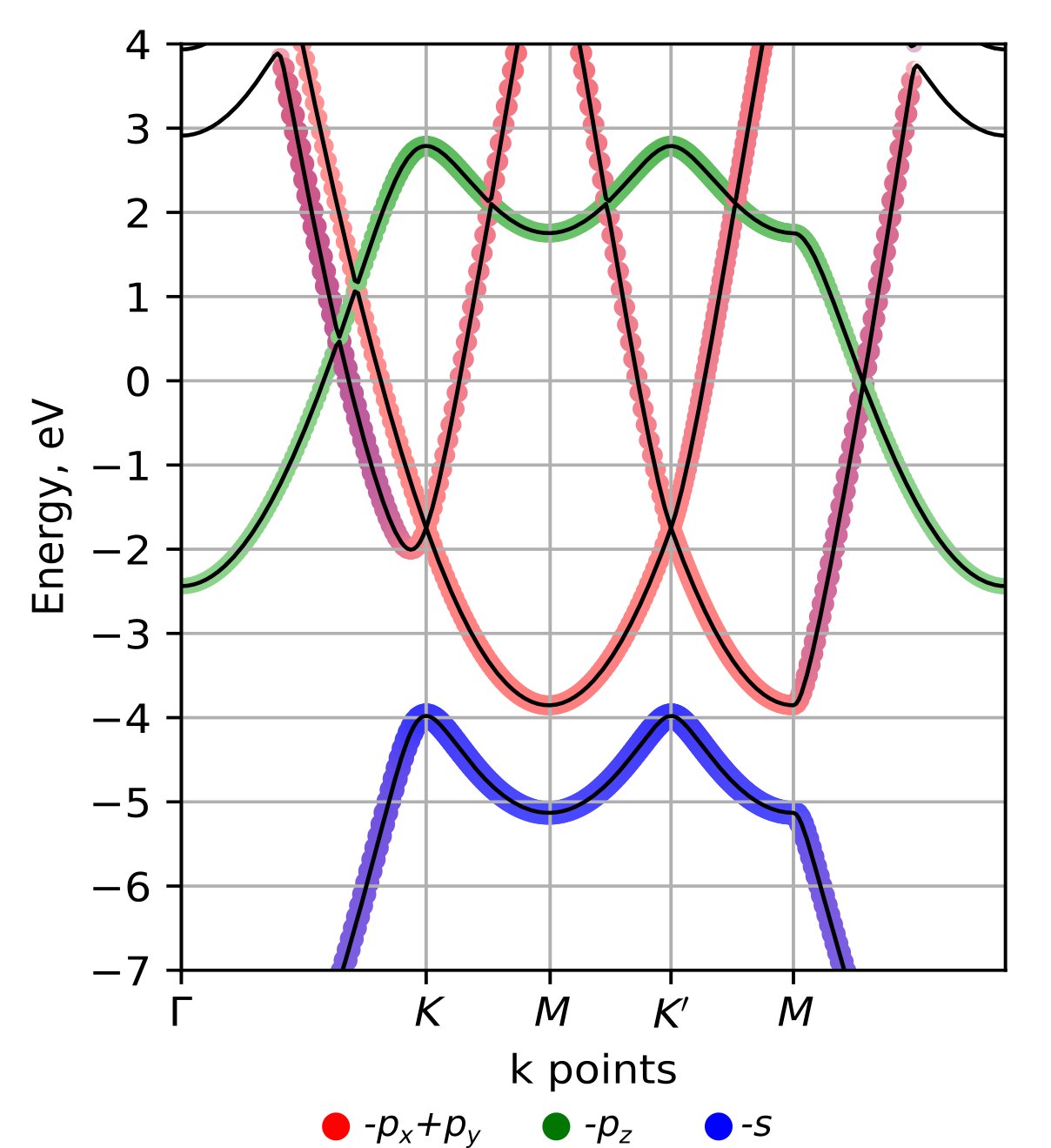}
    \caption{Band structure of standalone Sn triangular layer with orbital decomposition.}
    \label{fig:S1}
\end{figure*}

\begin{figure*}[ht]
    \centering
    \includegraphics[width=1.0\textwidth]{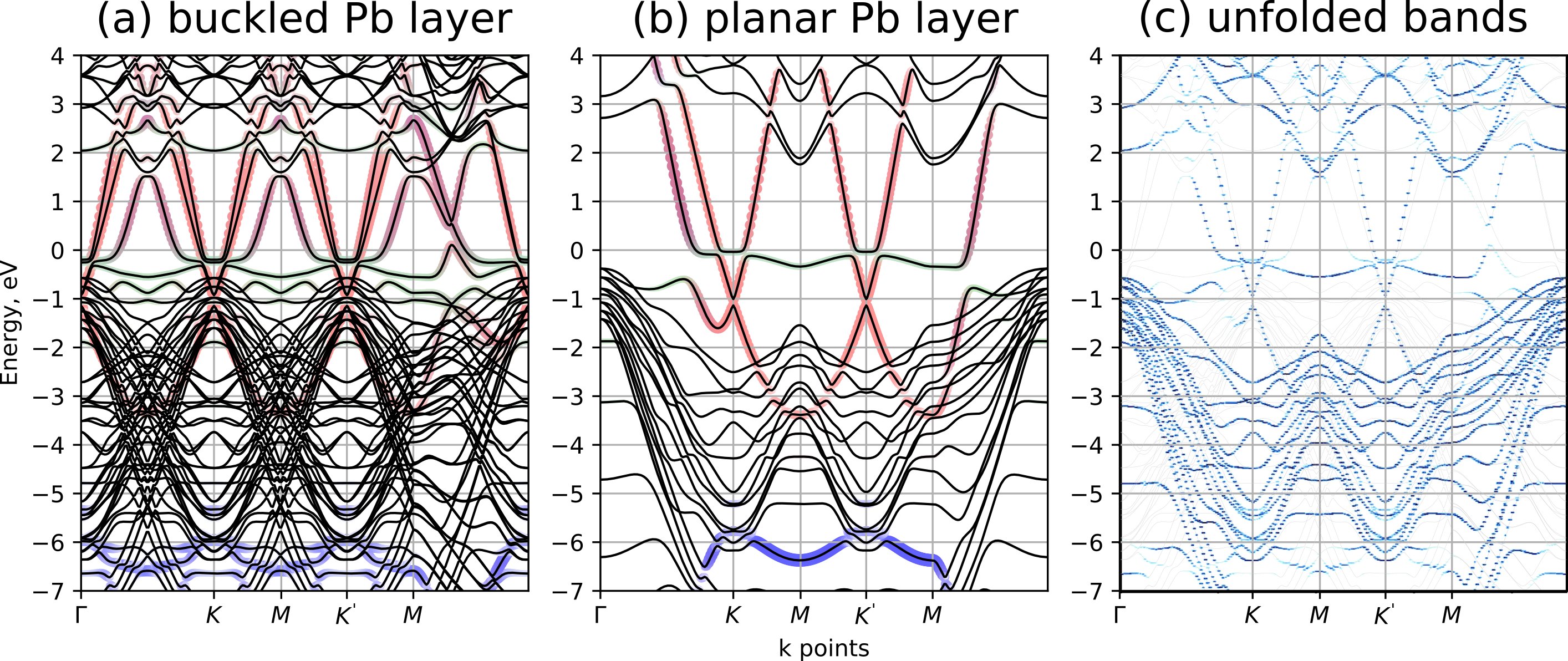}
    \caption{(a) Band structure of buckled $(\sqrt{3}\times\sqrt{3})$-Pb layer on SiC(0001) within $(1\times1)$ Brillouin zone. Colors show Pb orbital contribution as per orbital decomposition procedure: red -- $p_x+p_y$ orbital contribution, green -- $p_z$ contribution, and blue -- $s$ orbital contribution. (b) The same but for planar $(1\times1)$-Pb structure. (c) the result of band unfolding procedure. Lines show original folded band structure, while blue intensity map show weighted unfolded bands. One may see that the unfolded band structure is mostly identical to $(1\times1)$ planar band structure shown in (b).}
    \label{fig:S2}
\end{figure*}

\begin{figure*}[ht]
    \centering
    \includegraphics[width=1.0\textwidth]{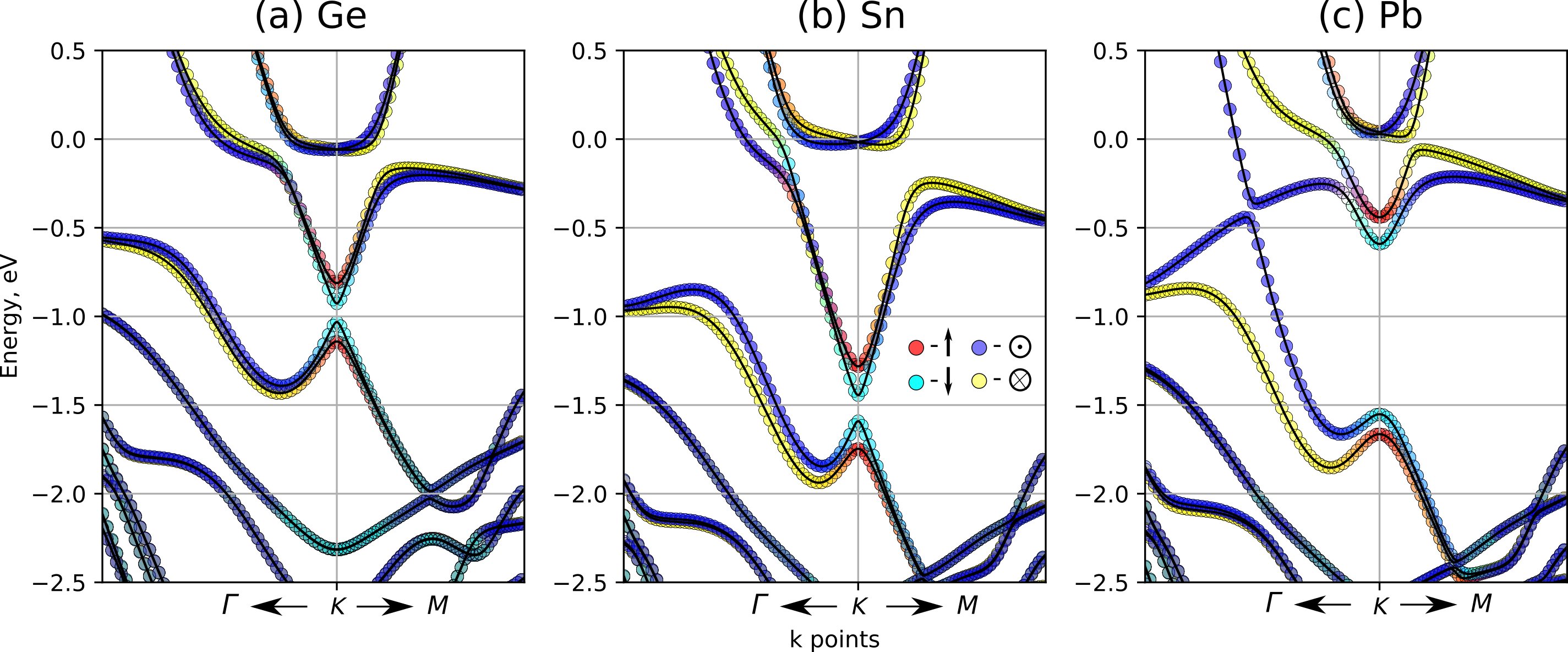}
    \caption{Band structure of Ge, Sn, and Pb layer using SOC in the vicinity of $\bar{K}$ point. Colors represent spin orientation. Note that bands near Fermi level show Rashba-like splitting in form of in-plane spin texture, while cone-like bands show Zeeman-like splitting with spin component normal to surface.}
    \label{fig:S3}
\end{figure*}

\begin{figure*}[ht]
    \centering
    \includegraphics[width=0.45\textwidth]{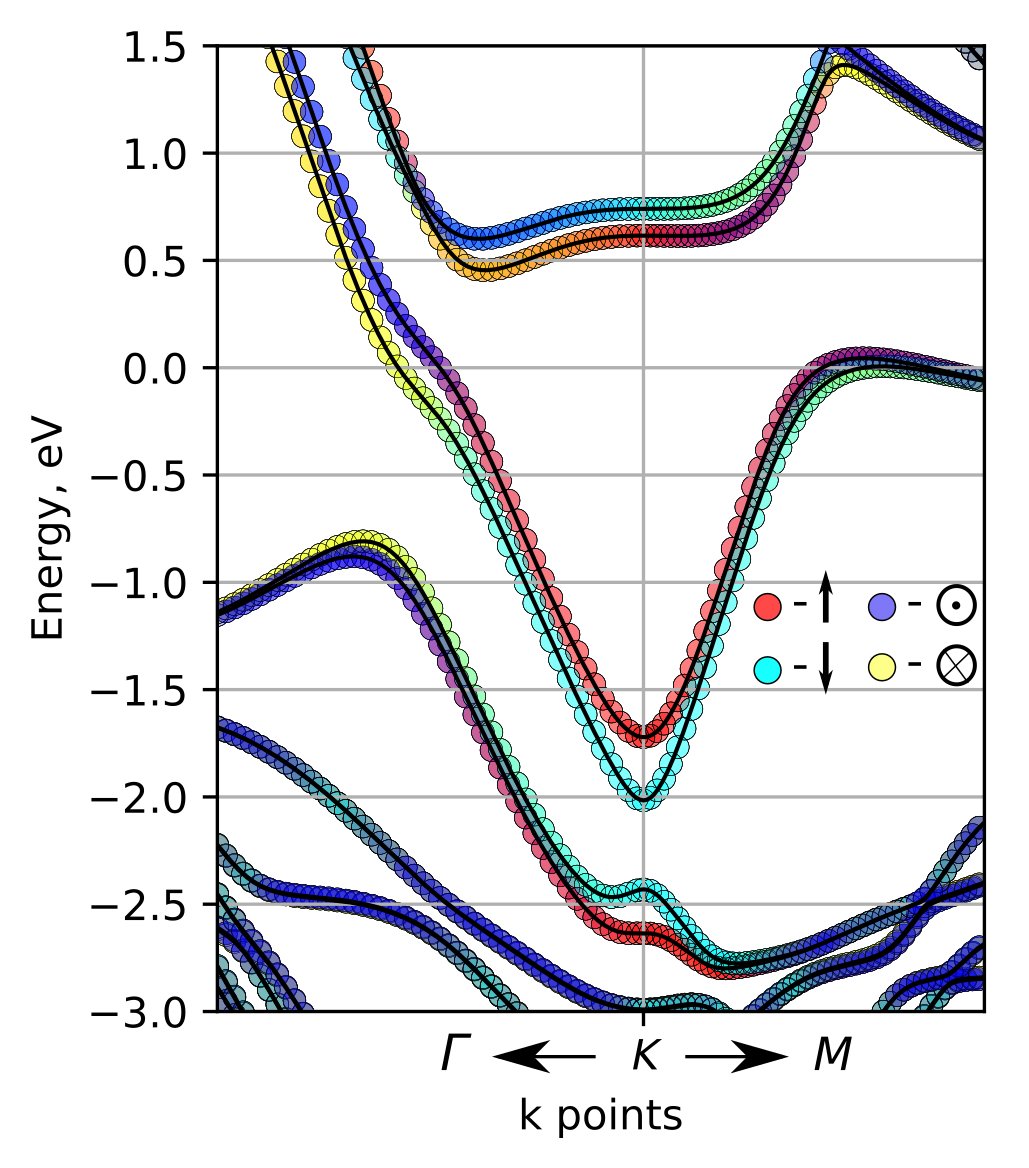}
    \caption{Band structure in the vicinity of $\bar{K}$ point of SiC(0001)-$(1\times1)$-Sn with SOC for the artificial case of Sn atoms occupying $T_4$ site. Note, that bands near Fermi level also exhibit Zeeman-like spin splitting. The band minima/maxima is still shifted from $\bar{K}$ point similar to the Rashba case.}
    \label{fig:S4}
\end{figure*}

\begin{figure*}[ht]
    \centering
    \includegraphics[width=0.6\textwidth]{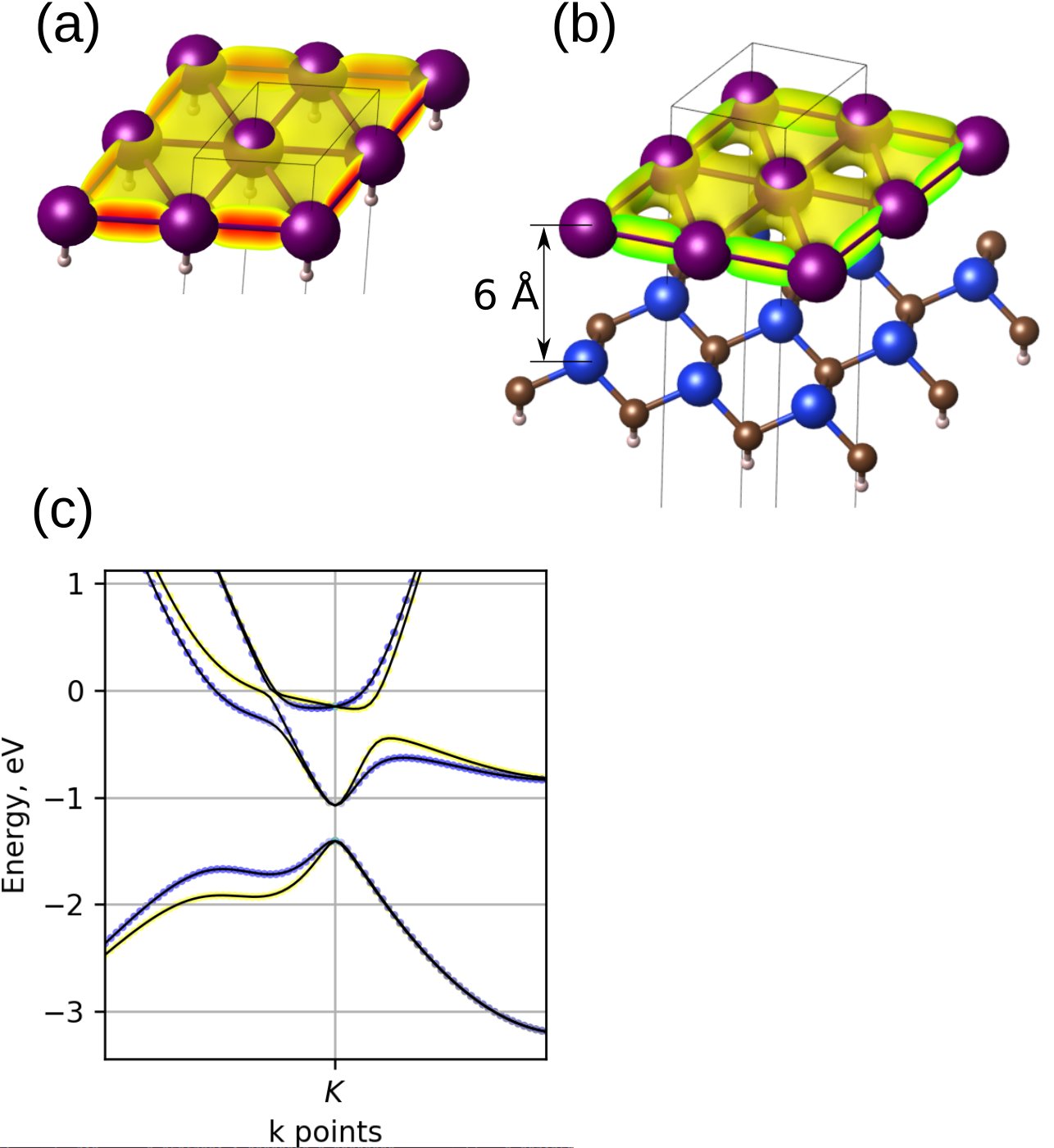}
    \caption{Partial charge density of one of $p_x+p_y$ states of TLAL layer. (a) Single-side hydrogenation introduce inversion symmetry breaking, while retaining system's $p6mm$ in-plane symmetry. The charge density is 6-fold symmetric, allowing Rashba-type splitting at $\bar{K}$ points, as evidenced from the band structure shown in (c). (b) TLAL layer on top of SiC bilayer. Although the interlayer distance artificially made rather large for direct interaction ($\sim6$~\AA{}), the substrate perturbation is enough to lower the symmetry of partial $p_x+p_y$ band charge density to $p3m1$ one, which forbids the Rashba-splitting at $\bar{K}$ points, but allow Zeeman-like splitting.}
    \label{fig:S5}
\end{figure*}

\end{document}